# Coexistence of Antiferromagnetic Spin Fluctuations and Superconductivity in La$_2$SmNi$_2$O$_7$ Thin Films

Minhui Xu,[1#] Yibo Wang,[1#] Jia Liu,[1] Long Cheng,[1] Shuyin Li,[1] Shuaishuai Yin,[1,2] Xu Zheng,[2] Lixin Yu,[1] Aidi Zhao,[1] Xiaolong Li,[2] Jiandi Zhang,[3,4†] Xiaofang Zhai[1,5†]

1 School of Physical Science and Technology, ShanghaiTech University, Pudong, Shanghai 201210, China

2 Shanghai Advanced Research Institute, Chinese Academy of Sciences, Pudong, Shanghai, 201204, China

3 Beijing National Laboratory for Condensed Matter Physics, Institute of Physics, Chinese Academy of Sciences, Beijing, 100190 China

4 College of Physics, University of Chinese Academy of Sciences, Beijing, 100190, China

5 State Key Laboratory of Quantum Functional Materials, School of Physical Science and Technology, ShanghaiTech University, Shanghai 201210, China

[#] *These authors contributed equally to this work.*

[†] *Corresponding authors: zhaixf@shanghaitech.edu.cn; Jiandiz@iphy.ac.cn.*

**Abstract**

The interplay between magnetic fluctuations and superconductivity is fundamental for understanding unconventional high-temperature superconductors. In the recently discovered Ruddlesden-Popper phase nickelates—which achieve superconducting transition temperatures up to ~100 K—this connection has been theoretically predicted but experimentally unverified. Using compressively strained $La_2SmNi_2O_7$ thin films, we report direct evidence for this interplay. We observe a characteristic 'Mexican hat'-shaped magnetoresistance, a signature of superconductivity coexisting with emergent antiferromagnetic (AFM) fluctuations. This distinctive feature arises from a competition between the magnetic field's suppression of AFM fluctuation-driven scattering and its suppression of superconducting fluctuations. We quantify these AFM fluctuations by a crossover field, $B^*$, whose magnitude decreases with increasing temperature and vanishes near the superconducting onset—behavior contrasting sharply with that in high-$T_c$ cuprates. Our results establish not merely coexistence, but also a direct and novel correlation between AFM instability and superconductivity in nickelates, offering crucial insight into their pairing mechanism.

**Introduction**

Recent studies have established the spin degree of freedom as a critical factor in the high-temperature superconductivity (SC) observed in bulk Ruddlesden–Popper (R-P) nickelates under high pressure, where it acts as both a competing order and a potential pairing mechanism [1,2]. At ambient pressure in the non-superconducting state, experiments reveal magnetic order in these materials. For instance, positive muon-spin relaxation finds magnetic order below 154 K in polycrystalline $La_3Ni_2O_{6.92}$ [3], while resonant inelastic X-ray scattering (RIXS) and nuclear magnetic resonance (NMR) detect a spin density wave (SDW) state in $La_3Ni_2O_7$ single crystals below ~150 K [4-6]. Under modest hydrostatic pressure (below ~2.3 GPa), the magnetic ordering temperature increases slightly, underscoring the robustness of these correlations [7]. However, upon further pressurization, the SDW state is suppressed well before the onset of SC. This has prompted a search for persistent magnetic fluctuations in the regime proximate to SC. Although theoretical work consistently identifies strong spin fluctuations as the likely pairing glue—leading to predicted sign-changing $s\pm$ or $d$-wave superconducting gaps [8-13]—direct experimental confirmation of magnetic correlations coexisting with or near the SC phase remains elusive, largely due to the formidable challenges of performing measurements under high pressure.

Motivated by the discovery of SC in high-pressure bulk phases, epitaxial thin-film growth has enabled superconducting $La_3Ni_2O_7$ and $La_{3-x}Pr_xNi_2O_7$ at ambient pressure [14,15], where compressive strain stabilizes SC with an onset $T_c$ around 40 K. Continued improvements in film quality have led to substantial progresses [16-18] and more recent work has demonstrated onset $T_c$ exceeding 60 K [19], suggesting that compressively strained thin films may ultimately achieve $T_c$ comparable to those of pressurized bulk materials. Benefitted from the ambient-pressure-stable SC thin films, advanced low-temperature electronic and spectroscopic probes have been applied, yielding important insights into the electronic structure and superconducting characters, including the $α$, $β$ and $γ$ pockets near the Fermi surface [20,21], the dimensionality of the SC [14,15,17] and the vortex state [19]. However, experimental signatures of SDW have not been observed in epitaxial films, suggesting its potential irrelevance to thin-film SC and a distinction from the high-pressure bulk SC phase. Consistent with this difference, theoretical studies predict enhanced spin fluctuations in strained $La_3Ni_2O_7$ films compared to bulk $La_3Ni_2O_7$ at ambient pressure [22,23], as well as a monotonic enhancement of interlayer AFM coupling under compressive strain [24]. Experimentally, evidences of SC coexisting with spin-glass state have been reported in $La_{2.46}Pr_{0.24}Sm_{0.3}Ni_2O_7$ and $La_{2.85}Pr_{0.15}Ni_2O_7$ films [25]. Together, these results highlight the strong potential for finding emergent magnetic correlations in superconducting R-P nickelate thin films under ambient pressure.

Here we report on our study of the magneto-transport behavior of superconducting $La_2SmNi_2O_7$ thin films which were fabricated under compressive strain on (0 0 1) $SrLaAlO_4$ (SLAO) substrates. X-ray diffraction (XRD) and synchrotron half-order diffraction confirm the films' high crystalline quality, verifying the characteristic layered 3-2-7 structure and an orthorhombic octahedral rotation pattern. Below the superconducting onset temperature, a pronounced *'Mexican hat'*–shaped magnetoresistance (MR) emerges when an in-plane magnetic field is applied. The center of this feature (low-field region, $|B| < B^*$, where $B^*$ is the crossover field at the resistance minimum) displays negative MR characteristic of AFM correlations, whereas the edge (high-field region, $|B| > B^*$) exhibits positive MR due to Zeeman pair-breaking of superconducting fluctuations. Notably, $B^*$—reflecting the strength of the underlying AFM correlations—decreases with increasing temperature and disappears as approaching to the onset of SC, thus different from that in high-$T_c$ cuprates. Furthermore, oxygen vacancies strongly suppress AFM fluctuations, as evidenced by diminishing $B^*$. These results demonstrate clear experimental evidence of AFM spin fluctuations coexisting with SC, indicating an exceptionally strong coupling between magnetism and SC in compressively strained $La_2SmNi_2O_7$ thin films under ambient pressure.

**Results**

We fabricated 10-nm-thick $La_2SmNi_2O_7$ films on (0 0 1) oriented SLAO substrates via pulsed laser deposition, capping them with a protective 1-unit-cell-thick layer of $SrTiO_3$ (STO). To compensate for oxygen deficiencies, the films were subsequently annealed in dilute ozone (see Methods in Supplementary Materials for details). Bulk $La_2SmNi_2O_7$ has a pseudo-tetragonal in-plane lattice constant of ~ 3.820 Å (Fig. 1a) [26,27]. The SLAO substrate, with a smaller in-plane lattice constant of 3.756 Å, therefore imposes a biaxial compressive strain on the film.

Structural characterization confirms the high quality of the strained films. The 2D

reciprocal space map (RSM) of the ozone-annealed film (Fig. 1b) shows it is fully coherent to the substrate, corresponding to a compressive strain of approximately 1.7%. XRD scans along the (0 0 $L$) and (1 0 $L$) directions, shown in Fig. 1c and Supplementary Materials (SM) Fig. S1 respectively, reveal reflections solely from the La$_2$SmNi$_2$O$_7$ phase, with a *c*-axis lattice constant of 20.53 Å, indicating high phase purity. We further investigated the oxygen octahedral rotation pattern using half-order diffraction measurements at beamline BL02U2 of the Shanghai Synchrotron Radiation Facility. As shown in Fig. 1d, we observe only ($n/2$, $m/2$, $l$) peaks—where $n$, $m$, and $l$ are integers relative to the pseudo-tetragonal reciprocal lattice ($k_a = k_b = 2\pi/3.756$ Å$^{-1}$, $k_c = 2\pi/20.53$ Å$^{-1}$)—while ($n/2$, $m$, $l/2$) and ($n$, $m/2$, $l/2$) peaks are absent. These observations indicate that the NiO$_6$ octahedra undergo out-of-phase rotations around *a*- and *b*-axes and in-phase rotations around the *c*-axis, corresponding to the Glazer notation of $a^-a^-c^+$.

Figure 1e displays the temperature-dependent resistivity ($\rho$) of an ozone-annealed La$_2$SmNi$_2$O$_7$ film (sample 1). The $\rho(T)$ curve reveals three distinct regimes: (i) above 68 K, the resistivity decreases with cooling, indicative of metallic behavior; (ii) between 21 K and 68 K, an upturn appears upon cooling, signaling an insulating tendency; and (iii) below 21 K, defined as the superconducting onset temperature $T_c^{onset}$, the resistivity drops sharply, marking the emergence of SC. This superconducting transition is further corroborated by the magnetic-field dependence of the resistivity, $\rho(B, T)$, shown in Fig. 2. The intermediate resistivity upturn resembles behavior observed in underdoped high-$T_c$ cuprates and has been attributed to Kondo scattering from defects such as oxygen vacancies [28]. A similar coexistence of a metal-insulator crossover and SC has been reported in oxygen-deficient (La, Pr)$_3$Ni$_2$O$_{7-\delta}$ thin films [29]. Consequently, residual oxygen vacancies likely persist in sample 1. This defect scenario may account for the absence of zero resistivity down to 0.3 K, as shown in the inset of Fig. 1e.

We further investigated the temperature-dependent resistivity under applied magnetic fields. Isomagnetic $\rho(T)$ curves were measured with fields oriented either perpendicular (Fig. 2a) or parallel (Fig. 2b) to the film plane. The superconducting onset temperature, $T_c^{onset}$, is strongly suppressed by a perpendicular field, while decreasing more gradually under a parallel field, highlighting a prominent anisotropy in the field response. The upper critical fields, $H_{c2\parallel}(T)$ and $H_{c2\perp}(T)$, were extracted for both field orientations using the 94% and 85% criteria of $\rho(T_c^{onset})$. Data for the 94% criterion (shown in Fig. 2a and 2b) and the 85% criterion (provided in SM Fig. S2) are plotted as functions of temperature in Fig. 2c. The resulting $H_{c2}(T)$ curves show excellent agreement with the 2D Ginzburg-Landau model, which is described by the following equations.

$$\mu_0 H_{c2\perp}(T) = \phi_0(1 - T/T_c)/[2\pi(\xi_{ab})^2], \qquad (1)$$

$$\mu_0 H_{c2\parallel}(T) = \phi_0\sqrt{12(1 - T/T_c)}/[2\pi\xi_{ab}d_{sc}], \qquad (2)$$

where $\phi_0$ is the quantum magnetic flux, $\xi_{ab}$ is the zero-temperature superconducting coherence length, and $d_{sc}$ is the effective superconducting thickness [30]. The upper critical fields at zero temperature exhibit significant anisotropy. For the out-of-plane direction, $H_{c2\perp}(0)$ values are 12 T and 11.5 T using the 94% and 85% resistance criteria, respectively. In contrast, the in-plane upper critical fields [$H_{c2\parallel}(0)$] are substantially larger, reaching 38 T and 39 T for the same criteria. From these values, the extracted coherence length ($\xi_{ab}$) and superconducting thickness ($d_{sc}$) are nearly identical, at 5.3–5.4 nm and 5.4–5.6 nm, respectively. The similarity between $\xi_{ab}$ and $d_{sc}$, which aligns with findings in prior (La,Pr)$_3$Ni$_2$O$_7$ thin film studies, suggests a quasi-2D superconducting state.

AFM spin fluctuations are revealed through in-plane MR measurements. As illustrated schematically in Fig. 3a, the resistivity displays a negative MR near zero in-plane field (forming the central peak of a 'hat'-shaped curve), independent of the in-plane field direction, while it transitions to a positive MR at higher fields (the edges of the 'hat'). To ensure accurate MR measurements, the magnetic field was carefully aligned parallel to the film plane by rotating the sample to locate the resistance minimum under a constant field (see SM, Methods and Fig. S3). Measurements were performed with the field aligned along the ***x***, ***y***, and diagonal directions, with current applied along ***x***. All configurations yielded similar results. As shown in Fig. 3b (***B // x***) and 3c (***B // y***), the resistivity $\rho$ was measured at a temperature of 2K under magnetic field sweeps from –9 T to 9 T and back to –9 T. A slightly asymmetric 'W'-shaped hysteresis loop is observed at both 2 K and 3 K. The two minima of $\rho$, marked by red dashed lines in Fig. 3c, define a crossover field $B^*$ of approximately 2.7 T.

The MR exhibits a field-induced crossover at $B^*$. For $|B| < B^*$, the resistivity decreases with field, producing a negative MR. In contrast, for $|B| > B^*$, it increases approximately as $\rho \propto |B|^2$. This positive $B^2$-dependence is attributed to Zeeman pair-breaking of superconducting fluctuations [31,32]

Above 5 K, below the superconducting transition, the 'W'-shaped $\rho(B)$ reverts to a conventional 'U'-shaped profile, confirming the disappearance of the zero-field peak. Furthermore, control measurements in the normal-state resistivity upturn regime (SM Fig. S4) show no negative MR, which rules out Kondo scattering as the origin of the low-field negative feature observed near $T_c$.

The resistivity peak at zero field excludes ferromagnetic (FM) contributions to the negative MR, as FM-dominated negative MR typically features as a butterfly shape near zero field [33]. Instead, a zero-field resistivity peak is a hallmark signature of AFM spin fluctuations, a feature previously observed in both electron- and hole-underdoped high-$T_c$ cuprates just above their superconducting onset temperature [28,34,35] and in $Sr_2IrO_4$ with an AFM ground state [36]. However, in underdoped cuprates, the negative MR associated with AFM fluctuations is confined to the resistivity upturn regime and disappears below $T_c^{onset}$. This behavior is markedly different from our observations in these compressively strained $La_2SmNi_2O_7$ thin films, where the AFM-type negative MR accompanies the SC below $T_c^{onset}$ and simultaneously enhances with SC.

Next, we exclude several other mechanisms that might account for the observed negative MR near zero in-plane field. First, the similarity in MR shapes for both $\boldsymbol{B}\!/\!/\boldsymbol{J}$ and $\boldsymbol{B}\perp\boldsymbol{J}$ orientations indicates that the Abrikosov-vortex flow effect, driven by the Lorentz force $\mathbf{F}\propto\boldsymbol{J}\times\boldsymbol{B}$, is not a significant contributor [37]. This mechanism is also known to produce a resistivity minimum at zero field [38], whereas we observe a pronounced resistivity peak at zero in-plane field. (We do observe a zero-field minimum when applying a perpendicular field, $B_z$, as shown in SM Fig. S5.) Second, dissipation related to Josephson vortices can be excluded, as its primary influence is on the *c*-axis resistivity [39], and our measurements are confined to the *ab*-plane. Third, while in-plane field suppression of magnetic fluctuations can induce negative MR near $T_c^{onset}$ in BCS-like superconductors such as Pb and LAO/STO [40], this effect typically persists both above and below $T_c^{onset}$, which fundamentally differs from our observation of a negative MR that appears exclusively below $T_c^{onset}$. Finally, we rule out potential thermal artifacts from eddy-current heating during field sweeps. By varying the sweep rate over a wide range (from 50 Oe/s to 150 Oe/s), we find that the hysteresis loops for both $B_x$ and $B_y$ remain unchanged (SM Fig. S6), confirming that eddy-current effects are negligible.

Figure 3d displays a series of minor hysteresis loops obtained by sweeping the in-plane field $B_y$ from –9 T to a successively decreasing maximum field, then back to –9 T. From these data, we identify two characteristic bounding fields: an upper field $B_{HF}$ ≈ 5 T, beyond which the positive-field hysteresis is fully suppressed, and a lower field $B_{LF}$ ≈ −2 T, where the negative-field hysteresis vanishes. Notably, $B_{LF}$ is close to the field $B^*$ of the local resistivity minimum, while $B_{HF}$ is near the closing field of the 'W'-shaped $\rho(B)$ curve. To account for this unusual hysteretic behavior, we propose a scenario where, for fields above $B_{HF}$, a significant fraction of Cooper pairs is broken, generating normal electrons with spin-glass-like correlations. These may then interact with the external field or with antiferromagnetically correlated quasiparticles to produce the observed loop features. A complete explanation, however, will require further theoretical investigation.

We further investigate a second La$_2$SmNiO$_7$ thin film (sample 2), which exhibits a slightly higher superconducting onset temperature ($T_c^{onset} \approx 24$ K) and no resistivity upturn above $T_c^{onset}$. XRD scans along the (0 0 $L$) and (1 0 $L$) directions (SM Fig. S7) confirm a pure 3-2-7 layered structure. Field-dependent resistivity measurements (Fig. 4a: ***B*** // ***z***; Fig. 4b: ***B*** // ***x*** // ***J***) reveal a similar quasi-2D character, with corresponding Landau–Ginzburg model fits for the upper critical fields provided in SM Fig. S8. As shown in Fig. 4c, the in-plane MR (with ***B*** aligned along a diagonal direction) displays a pronounced 'W'-shaped profile from 2 K to 23 K. Compared to the sample 1, the sample 2 exhibits a substantially broader 'W' feature, with the local minimum field $B^*$ reaching ~ 8 T at 2 K. As shown in Fig. 4d, $B^*$ increases monotonically with decreasing temperature and shows no saturation down to 2 K. Upon approaching $T_c^{onset}$, $B^*$ decreases continuously to zero and is no longer observable above $T_c^{onset}$.

AFM spin fluctuations emerge below the superconducting onset $T_c^{onset}$, establishing R-P phase nickelates as fundamentally distinct from cuprates and iron pnictides. This difference potentially originates from the 3-2-7 nickelates hosting both $d_{x^2-y^2}$ and $d_{3z^2-r^2}$ orbitals near the Fermi level, unlike cuprates with a single active $d_{x^2-y^2}$ orbital. Although chemically unstable, interlayer apical oxygens are essential for mediating the $d_{3z^2-r^2}$ interlayer coupling and stabilizing the proposed $s \pm$ superconducting state [41], making suppression of oxygen vacancies critical. This is consistent with the results from inelastic neutron scattering (INS) [42] and RIXS [4] experiments, as well as the recent theoretical analysis [43]. Accordingly, we observe stronger AFM spin fluctuations in samples with fewer oxygen vacancies (sample 2 compared to sample 1), and the simultaneous enhancement of AFM fluctuations and superconducting correlations upon cooling points to an unconventional pairing mechanism unique to the 3-2-7 phase nickelates.

Finally, we examine the angular dependence of the resistivity and the characteristic field $B^*$ in the sample 2 at 2 K, with results presented in Figs. 4e and 4f. A negative MR emerges only when the magnetic field is aligned nearly within the film plane. Here, $\theta$ denotes the angle between the field and the sample normal (inset of Fig. 4b). As shown in Fig. 4e, the negative MR is clearly observed for $\theta$ between 90° and 91.9°. For larger angles ($\theta > 91.9°$), the negative MR vanishes and a positive MR dominates. We attribute this behavior to the growing perpendicular field component $B\cos\theta$, which becomes sufficiently large to generate a substantial vortex density that strongly enhances dissipation and raises the resistivity.

Figure 4f displays the angular dependence of the AFM-related crossover field $B^*$. At perfect in-plane alignment ($\theta = 90°$), $B^*$ attains its maximum value of 9 T. As $\theta$ deviates from 90°, $B^*$ decreases continuously, falling to 4.34 T at $\theta = 91.6°$. For further deviations, the negative MR signal becomes indistinguishable from the dominant positive background, preventing a reliable determination of $B^*$. Consequently, the extracted value of $B^*$ is highly sensitive to the precision of the in-plane field alignment.

**Conclusion**

In summary, we have established that the superconducting state in compressively strained $La_2SmNi_2O_7$ thin films is intrinsically linked to strong AFM spin fluctuations. This conclusion is anchored by the observation of a characteristic 'Mexican hat'-shaped

MR below the superconducting onset temperature, which uniquely combines an AFM-type negative MR at low fields ($|B| < B^*$) and a positive MR from superconducting pair breaking at high fields ($|B| > B^*$). Crucially, these AFM spin fluctuations develop *within* the superconducting state —a distinct behavior that sets this nickelate apart from cuprates. The enhancement of the crossover field $B^*$ with improved sample quality (reduced defects) and with decreasing temperature confirms the intrinsic nature of this interplay. Collectively, our results demonstrate that AFM spin fluctuations are fundamental to the superconducting mechanism in this R-P phase nickelate, pointing toward a multi-orbital SC characterized by a strong, cooperative relationship between magnetism and SC.

**Acknowledgement**

The work was financially supported by National Key R&D Program of China (Nos. 2022YFA1403000, 2023YFA1406301). The research used resources from Analytical Instrumentation Center (#SPST-AIC10112914) and Soft Matter Nanofab (SMN180827) in ShanghaiTech University. We thank the staff at beamline BL02U2 of the Shanghai Synchrotron Radiation Facility.

**Data availability**

The data that support the findings of this work are available from the authors upon reasonable request.

**Figures**

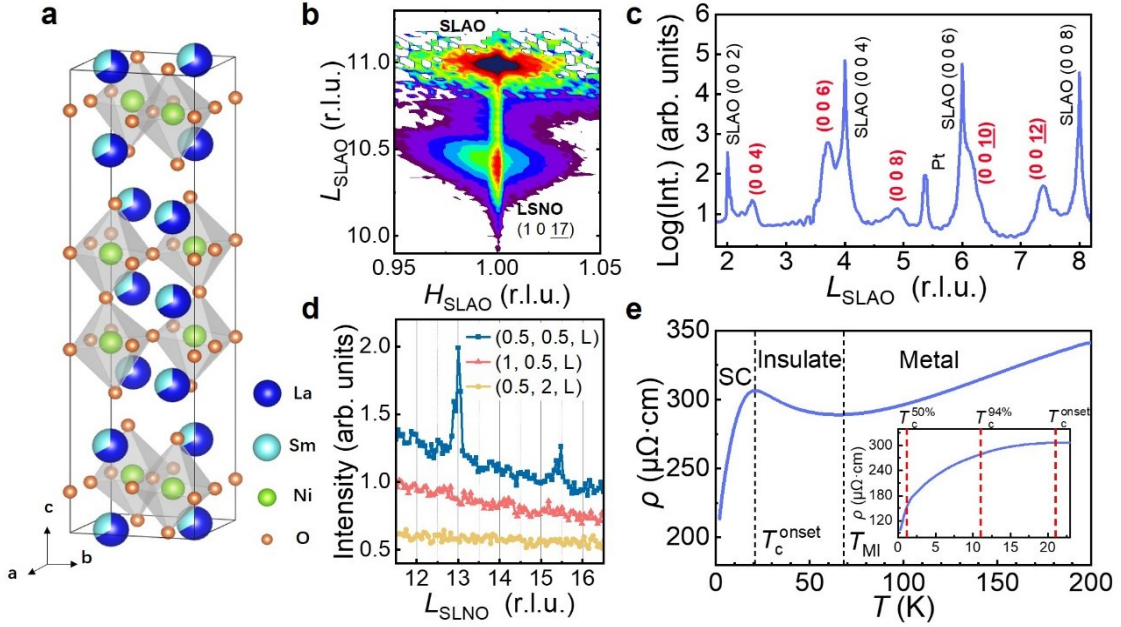

**Figure 1. The structural characterization and SC transition in resistivity measurements. a.** The schematic lattice structure of the La$_2$SmNi$_2$O$_7$ (LSNO) thin film. **b.** The reciprocal space map of the film (1 0 $\underline{17}$) peak and the SLAO substrate (1 0 $\underline{11}$) peak. **c.** The (0 0 $L$) diffraction scan of the film diffraction peaks, indicated by red color indices. The Pt peak is from the top electrode used for transport measurements. **d.** The synchrotron half order measurements along three characteristic scan directions. **e.** The superconducting transition measured from 200 K to 2 K. The inset show the transition from 25 K to 0.3 K measured in a $^4$He/$^3$He cooling environment.

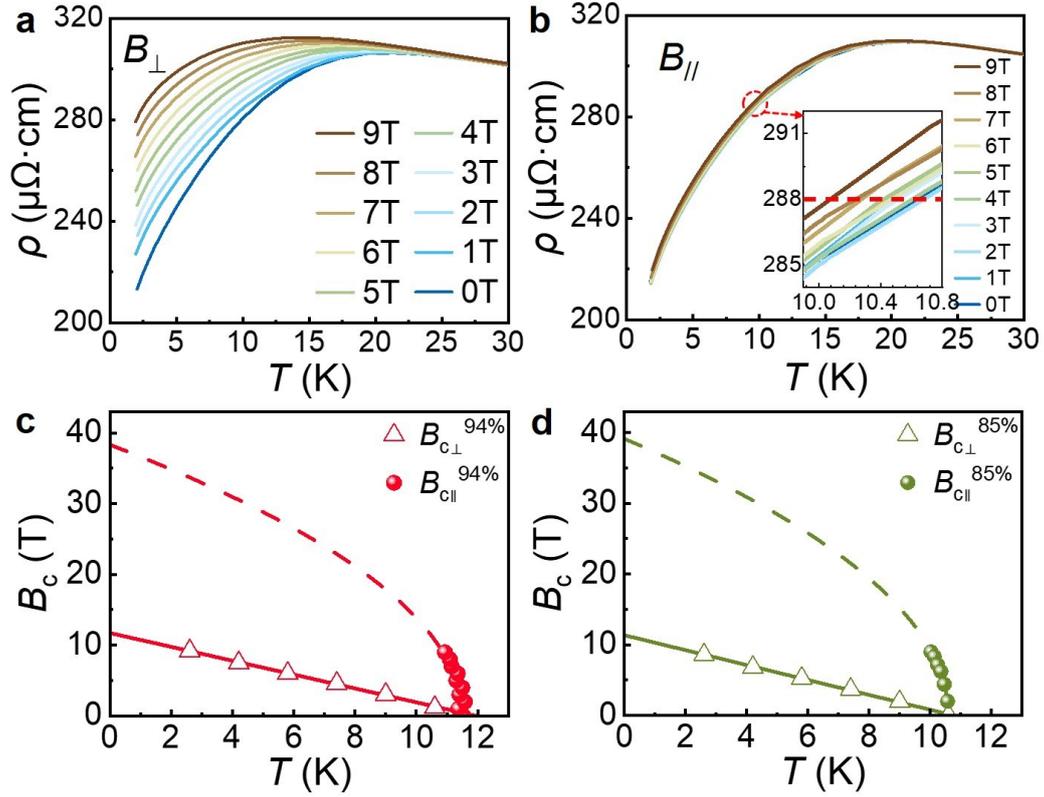

**Figure 2. The field dependence of the SC transition and the Landau-Ginzburg (L-G) model fitting. a.** The perpendicular field dependence of the SC transition revealing the onset $T_c$ is 24 K. **b.** The weak dependence of the SC transition on the parallel field. The inset shows the enlarged view of the resistivity which is about 94% (indicated by the red dash line) of the onset-temperature resistivity. **c.** The L-G model fitting using the field that reduces the normal state resistivity to 94% and 85% as the critical field. The extraction of the 94% set of data is done using sweeping temperature data from **a** and **b**, while the 85% set of data is extracted from sweeping field measurements shown in SM Fig. S2.

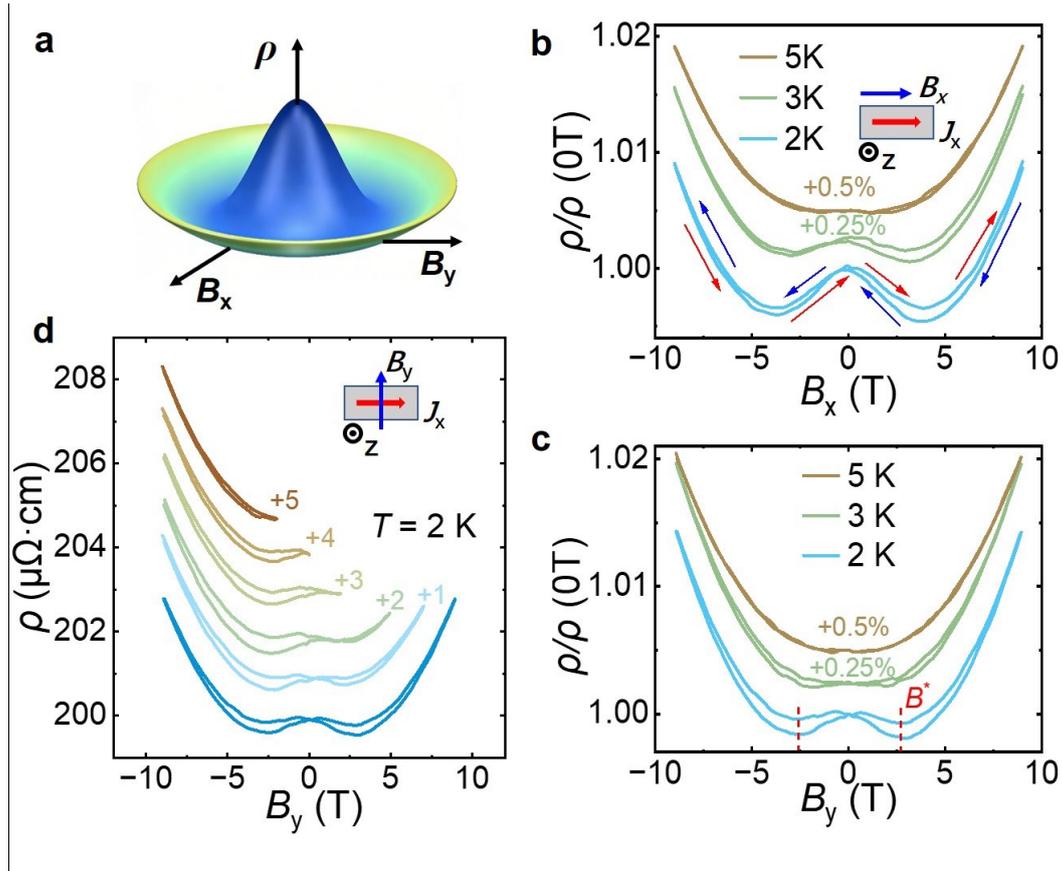

**Figure 3. The AFM spin fluctuation revealed by in-plane MR. a.** The 'Mexican hat' cartoon that schematically shows the dependence of resistivity as a function of in-plane field. The central region of the 'hat' indicates negative MR while the edge of 'hat' indicates positive MR. **b. c.** The reduced resistivity $\rho/\rho(0T)$ measured with in-plane magnetic field along *x* and *y* directions at 2 K, 3 K and 5 K. The 3 K and 5 K data are stacked up by adding 0.25% and 0.5% respectively. In **b**, red arrows indicate the −9 T to 9 T sweeping and blue arrows indicate the 9 T to −9 T sweeping. Red dash lines in **c** indicate the crossover field of $B^*$. **d.** The minor loops measured at 2 K with field aligned along the y direction. The loops are stacked by adding 1 μΩ·cm for each curve. The measurement geometries are shown in the insets of **b** and **d.**

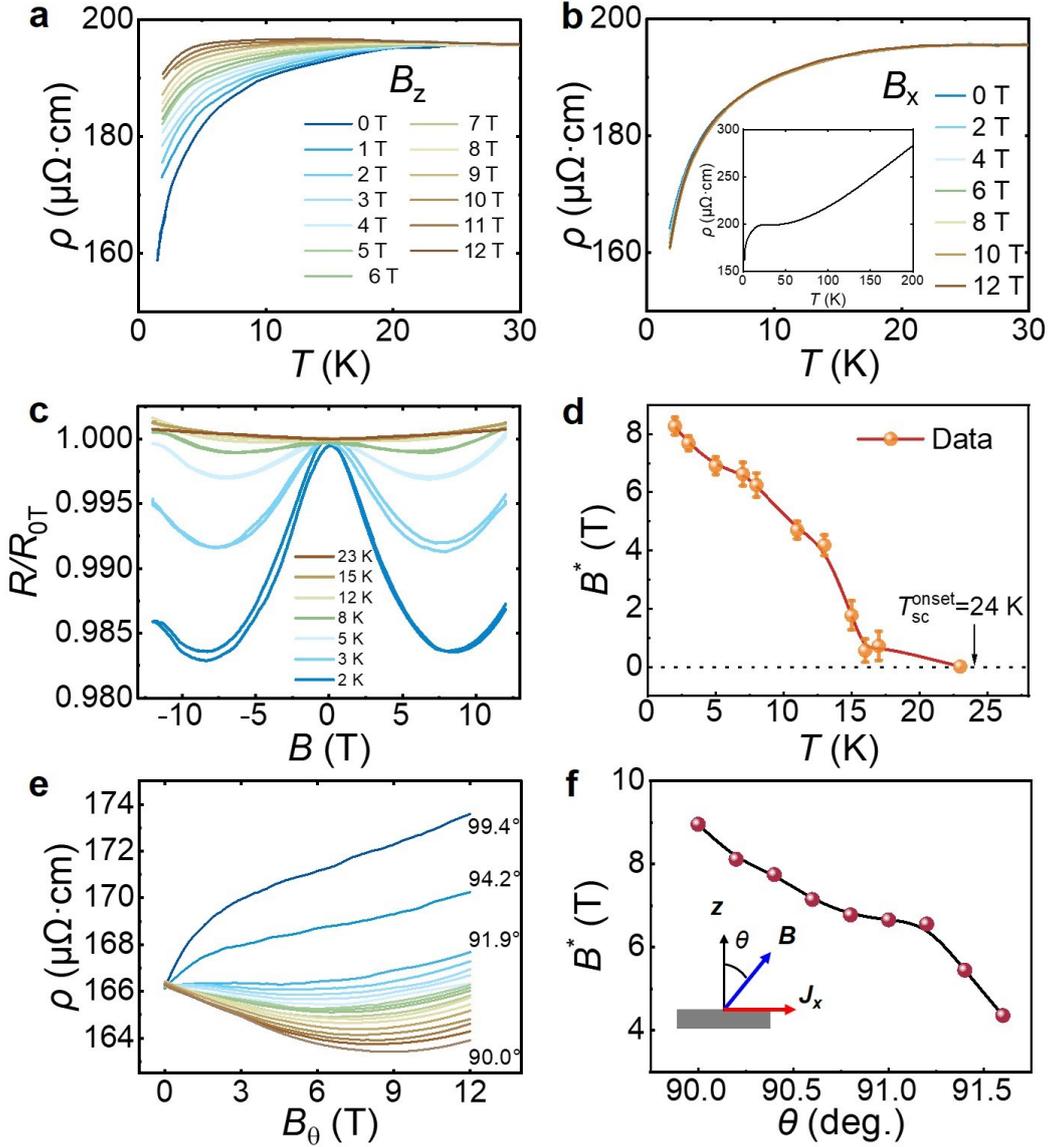

**Figure 4. The enhanced AFM spin fluctuation in a second sample with higher SC $T_c^{onset}$ and absence of resistivity upturn. a.** The SC transition measured with magnetic field perpendicular, and **b.** parallel to the film plane (***B***//***J***//***x***). The inset in **b** shows the resistivity in the entire temperature range from 2 K to 200 K. **c.** The reduced resistivity measured by sweeping the in-plane (along the *xy* diagonal direction) field between −12 T and 12 T at 2, 3, 5, 8, 12, 15 and 23 K. **d.** The dependence of the AFM characteristic field $B^*$ on temperature. **e.** The resistivity measured at 2 K with field rotating in the *x*-*z* plane from the parallel direction ($\theta = 90°$) to slightly off-parallel ($\theta = 99.4°$). **f.** The angular dependence of $B^*$ on $\theta$ obtained from **e**. The inset in **f** shows the measurement geometry of **e.**